\documentclass[prb,reprint,twocolumn]{revtex4}
\usepackage{graphicx} % Required for including images
\usepackage{amsmath} % For typesetting math
\usepackage{amssymb} % Adds new symbols to be used in math mode

\usepackage{color}
\begin{document}
\title{Interplay of Purcell effect, stimulated emission, and leaky modes in photoluminescence spectra of microsphere cavities}
\author{Ching-Hang Chien$^{1,2,3}$}
\author{Shang-Hsuan Wu$^{1,4}$}
\author{Trong Huynh-Buu Ngo$^{1,5}$}
\author{Yia-Chung Chang$^{1}$}
\email[Corresponding author. E-mail:] {yiachang@gate.sinica.edu.tw}
\address{$^{1}$ Research Center for Applied Sciences, Academia Sinica, Taipei 115, Taiwan}
\address{$^{2}$ Nano Science and Technology Program, TIGP, Academia Sinica, Taipei 115, Taiwan}
\address{$^{3}$ Department of Engineering and System Science, National Tsing Hua University, Hsinchu 300, Taiwan}
\address{$^{4}$ Department of Electrical and Computer Engineering, University of Texas at Austin, USA}
\address{$^{5}$ Department of Electrical and Computer Engineering, University of Alberta, Edmonton, Alberta T6G 2V4, Canada}

%\affiliation{Research Center for Applied Sciences, Academia Sinica, Taipei 115, Taiwan}

\date{\today}% It is always \today, today,

%%%%%%%%%%%%%%%%%%%%%%%%%%%%%%%%%%%%%%%%%%%%%%%%%%%%%%%
%\textbf{OUTLINE}
%%%%%%%%%%%%%%%%%%%%%%%%%%%%%%%%%%%%%%%%%%%%%%%%%%%%%%%

\begin{abstract}
A theoretical model for describing the emission spectra of microsphere cavities is presented, and its predictions of detailed lineshapes of emission spectra associated with whispering gallery modes (WGMs) of various orders in ZnO microspheres (MSs) are verified experimentally by photoluminescence (PL) spectroscopy.  The interplay of Purcell effect, quality factor, and leaky modes in spontaneous and stimulated emission spectra related to WGMs of all orders is revealed. The key success of the theory is based on the expansion of the full Green function of the MS in terms of all possible resonance modes in complex frequency space, which allows incorporation of contributions from leaky modes, stimulated emission processes, and Purcell effect.  We show that the spontaneous emission spectrum calculated according to Mie theory (without Purcell effect) is dominated by the contribution of leaky modes, while the spontaneous and stimulated emission enhanced by Purcell effect are responsible for the main WGM resonance peaks observed experimentally. It is found that the stimulated emission peaks are doubly enhanced by their respective mode quality factor Q: one factor from the Purcell effect and the other factor from the photon number derived from the rate equation. After combining all these effects  the theory can provide a quantitative description of fine features of both TE and TM modes (including higher-order modes) observed in the PL spectra of ZnO MSs. Surprisingly, it is found that for ZnO MS with diameter larger than 5 $\mu m$, the PL emission spectrum is dominated by higher-order modes. The quantitative understanding of the interplay of these emission mechanisms should prove useful for optimizing the performance of light-emitting devices based on micro resonators.

\end{abstract}
%Use showkeys class option if keyword
\maketitle

%\section{Introduction}

There has been a growing interest in the study of optical emission related to resonant modes in micro cavities with high quality ($Q$) factor due to their potential applications in light emitting, optical amplification, sensing, and modulation\cite{App1,App2,App3,App4}.  The resonant modes in cavities with round surface (e.g. spheres, domes, discs,  and cylinders) are usually called  whispering gallery modes (WGMs).
In multi-faceted cavities, the resonance frequencies of these WGM modes can be determined by geometric optics\cite{Geo1,Geo2}, if the cavity size is large compared to wavelength. When the cavity size is comparable to wavelength, especially for high-symmetry shape (e.g. spherical), a complete description of the resonant frequency and $Q$ factor of WGMs must require wave mechanics.

Despite numerous efforts, there remains a wide gap between wave-mechanical simulations and  experimental emission spectra. Little\cite{Little} discussed the relation between the $Q$ factor and energy loss of a microsphere coupled to a fiber. Schiller \cite{Schiller,Schiller2} proposed a method to solve efficiently the characteristic equation for WGMs of microspheres for large mode number. The WGM frequencies are known to be sensitive to the porosity inside the microsphere and roughness on the surface. Yet to the best of our knowledge, a theory capable of predicting  peak positions and relative intensities of WGMs well enough to identify all fine features of observed emission spectra is still lacking. Three mechanisms are incorporated in this work. The first is the contribution from the leaky modes\cite{10,11}.  The second is the enhancement due to Purcell effect\cite{Purcell} for high-Q WGM peaks\cite{13,14,15,16}. For this, we derived an simple analytic expression of the Purcell factor for a microsphere (MS). Finally, we combine both spontaneous and stimulated emission processes\cite{17,Einstein} to provide an explanation of detailed lineshapes of emission spectra from ZnO microsphes of various sizes.

For any two level system with an occupied  upper level, there is spontaneous emission as long as the optical coupling  between them is significant enough to overcome the nonradiative emission. Such a two-level system can be called a light emitter. Moreover, in the presence of electromagnetic (EM) field, stimulated emission\cite{Einstein} can also occur. For a system consisting of a large number of light emitters, the population of electrons in the upper and lower levels of these emitters can be determined by solving the rate equations\cite{book0,book1,book2}, which requires knowledge of the feeding rate, the radiative and nonradiative decay rate, and the optical pumping rate for each level.

The spontaneous emission rate depends on the photon mode density of states (DOS), which can be rather sensitive to the emission frequency in a resonant cavity, as the mode DOS is sharply concentrated at the resonance frequency of a cavity. This leads to an enhancement of spontaneous emission rate, a phenomenon called the Purcell effect\cite{Purcell}. The Purcell effect is known to play a crucial role in the understanding of the emission spectra in a cavity\cite{PB,PF,15,16}. The Purcell effect beautifully shows that by carefully designing the cavity environment where a light source is embedded, the spontaneous emission rate of the emitter {\it at resonance} can increase dramatically by a Purcell factor given by\cite{Purcell}
$P_F=\frac{3}{4\pi^2}\frac{\lambda_r^3 Q}{V_{\mu}}$,
where $\lambda_r=\lambda/n_r$ is the light wavelength inside the cavity with refractive index $n_r$; $Q$ is the quality factor,  and $V_{\mu}$  the modal volume of the cavity\cite{Vmu}. Purcell effect in acoustic cavities also attracts significant interest recently\cite{Ying}.

Furthermore, under continuous optical pumping, the photon number generated by a given emitter for mode $i$ (modeled by a  two-level system) obeys the rate equation\cite{PF}
\(
\frac{\partial n_i}{\partial t} = R_{sp}(\omega_i)+ \gamma_i g_i n_i  - n_i/\tau_i,
\)
where $R_{sp}(\omega_i)=\gamma_i f^U_i(1-f^L_i)P_F(\omega_i)$ is spontaneous emission rate for the two-level system and $g_i=f^U_i-f^L_i$ denotes the distribution function of the population inversion with $f^U_i (f^L_i)$ representing the carrier population in the upper (lower) level. $\gamma_i$ is the corresponding recombination rate, $n_i$ is the photon number generated by an emitter, and $\tau_i$ is the photon lifetime of the cavity mode $\omega_i$.

In steady state, the photon number per emitter is given by $n_i= \tau_i R_{sp}(\omega_i)/(1-\tau_i \gamma_i g_i)$, while in a cavity, the photon lifetime $\tau_i\equiv \tau_0 Q_i$ is  proportional to the quality factor $Q_i$, as can be understood that  the higher the quality factor, the longer the photon lives. Here $\tau_0$ is a characteristic time scale. So the photon number with mode frequency $\omega_i$ increases with $Q_i$. Due to the unavoidable surface roughness and other external scattering mechanisms, the effective $Q_i$ may be reduced. We may write $Q_i^{-1}=Q_{0i}^{-1}+Q_{ex}^{-1}$, where $Q_{0i}$ denotes the $Q$ factor of mode $i$ in an ideal resonator and $Q_{ex}$ the limiting $Q$ factor due to external scattering mechanisms\cite{Qext}.

Thus for mode $i$, the corresponding stimulated emission rate
%\begin{equation}
\( R_{st}(\omega_i)\propto n_i g_i\gamma_i \) .
%\end{equation}
is then enhanced again by the quality factor $Q_i$.
All this shows that the photon emission spectrum will be doubly enhanced by the quality factor $Q_i$ for the stimulated emission, which dominates for high-$Q$ modes. For low-$Q$ modes, the spontaneous emission can still dominate in the PL spectrum, even though it is enhanced by the $Q$ factor only once.  Combining both the stimulated and spontaneous emission processes is crucial for correctly accounting for the relative peak intensities in the PL emission of microcavity for TE/TM modes of various orders.

For some micro cavities, it is found that the leaky modes also have significant contribution to the PL emission spectra\cite{10}. To correctly describe the relative contribution of the leaky modes and resonance modes, one needs to derive the full Green function (GF), $G({\bf r},{\bf r'})$ of the micro resonator, which describes the EM field detected at point ${\bf r'}$ caused by a point source of unit strength at ${\bf r}$. In this paper, we first derived  the full GF of the MS and its Purcell factor. Then, we devised a scheme to find the relations between contributions of leaky modes, spontaneous, and stimulated emission to the PL spectra of spherical micro cavities. We also fabricated a ZnO MS with diameter around 5 $\mu m$ and performed PL measurements to compare with predictions of the theory. This large-size ZnO MS allows us to identify contributions from many higher-order modes and provide better understanding of the physical mechanisms behind.

For synthesizing the ZnO microspheres, a hydrothermal method was used based on our previous report \cite{Rakesh,Trong}. In short, aqueous solutions of 50 mM zinc nitrate hexahydrate (Zn(NO$_3)_2 \cdot $6H$_2$O), 50 mM zinc hexamethylenetetramine (HMT, C$_6$H$_{12}$N$_4$), and 35 mM trisodium citrate were first prepared. Next, 10 ml of zinc nitrate solution was mixed with 10 ml of HMT and 10 ml trisodium citrate solutions. The mixture solution was stirred in the closed glass bottle thoroughly and transferred to oven for hydrothermal growth at 90$^0C$ for 90 min.
The as-prepared ZnO microspheres/ethanol solution was dropped to precleaned Si substrate and then annealed at 550$^0C$  for 12 hours in air. The surface morphology and size of ZnO microspheres were measured by a scanning electron microscope (SEM-Nano Nova). The PL spectra of ZnO microcavities of various sizes was obtained using a micro PL system (Horiba Jobin Yvon HR-800) with a 325 nm He-Cd CW laser as the excitation source and a 2400 grooves/mm grating in the backscattering geometry.

%For ZnO  micro cavity considered here,  part of the light emission is related to defect states with emission spectra spread over the entire visible range.

%\section{Green Function of a microsphere}

Consider a point source field ${{\bf f}_\alpha }\left( {{\bf r}'',{{\bf r}_j}} \right)$ at ${\bf r}_j$ (with  polarization along $\alpha$ axis),  the induced field at ${\bf r}'$ is related to the {\it full} GF ${{\mathord{\buildrel{\lower3pt\hbox{$\scriptscriptstyle\leftrightarrow$}}\over
 G} }}$   as\cite{PB}
\begin{equation}
{\bold E}( {k_0,{\bf{r}}',{\bf r}_j}) = {\int d{{\bf r}''{{\mathord{\buildrel{\lower3pt\hbox{$\scriptscriptstyle\leftrightarrow$}}\over
 G} }}( k_0,{\bf{r}}',{\bf{r}}'') \cdot {\bf{f}_\alpha }\left( {{\bf r}'',{{\bf r}_j}} \right)} }\,.
\end{equation}
Here $k_0=\omega/c$ is the  photon wave number in free space.
The full GF of a MS obeys the Dyson equation\cite{GF}
\begin{equation}\label{eq:fullGF}
    \begin{aligned}
&{\mathord{\buildrel{\lower3pt\hbox{$\scriptscriptstyle\leftrightarrow$}}\over
 G} }\left( {k_0,{\bf{r}},{\bf{r'}}} \right)  = \mathord{\buildrel{\lower3pt\hbox{$\scriptscriptstyle\leftrightarrow$}}\over
 G} _0\left( {{\bf r},{\bf{r'}}} \right)
 &\\
  &+ k_0^2\left( {{\varepsilon_1} - 1} \right)\int_0^R {{{\tilde r}^2}d\tilde r} \int {d\tilde \Omega } {\mathord{\buildrel{\lower3pt\hbox{$\scriptscriptstyle\leftrightarrow$}}\over
 G} _0}\left( {{\bf r},{\bf{\tilde r}}} \right) \cdot {\mathord{\buildrel{\lower3pt\hbox{$\scriptscriptstyle\leftrightarrow$}}\over
 G}}\left( {{\bf{\tilde r}},{\bf{r'}}} \right) &
        \end{aligned}
\end{equation}
where $\mathord{\buildrel{\lower3pt\hbox{$\scriptscriptstyle\leftrightarrow$}}\over
 G}_0\left( {{\bf r},{\bf{r'}}} \right)$ is the unperturbed GF in the absence of the ZnO microsphere with a dielectric constant $\varepsilon_1$. We expand the unperturbed GF in terms of  Mie basis functions $\{{\bf{M}}_{\ell,m}^{<,>}\left( {k_0{\bf{r}}} \right),{\bf{N}}_{\ell,m}^{<,>}\left( {k_0{\bf{r}}} \right)\}$ as \cite{GF,Mie,res,BH,SM}
\begin{eqnarray}
  && \mathord{\buildrel{\lower3pt\hbox{$\scriptscriptstyle\leftrightarrow$}}\over
 {G_0}}\left( {{\bf r},{\bf{r'}}} \right) =  - {1 \over {k_0^2}}\delta \left( {{\bf r} - {\bf{r'}}} \right){{\bf{e}}_r}{{\bf{e}}_r} + ik_0 \sum\limits_{\ell,m}   \\
  &&
\left[ {\bf{M}}_{\ell, m}^{<,>} \left( {k_0{\bf{r}}} \right)\bar{\bf{M}}_{\ell, m}^{>,<} \left( {k_0{\bf{r'}}} \right) + {\bf{N}}_{\ell,m}^{<,>} \left( {k_0{\bf{r}}} \right)\bar{\bf{N}}_{\ell m}^{>,<} \left( {k_0{\bf{r'}}} \right) \right] \nonumber.
\end{eqnarray}
In the above, the 1st (2nd) superscript of $\{{\bf{M}}_{\ell,m},{\bf{N}}_{\ell,m}\}$ is used when $r<r'$ ($r>r'$).
The full GF can be expressed as a sum of TE and TM contributions with\cite{SM}
 \begin{equation}\label{TE+TMGF}
    \begin{aligned}
    &{\mathord{\buildrel{\lower3pt\hbox{$\scriptscriptstyle\leftrightarrow$}}\over
 G}_{TE}}\left( {k_0,{\bf{r}},{\bf{r'}}} \right) = \sum\limits_{\ell,m} {{c_\ell }{\bf{M}}_{\ell,m}^ < \left( {{k_1}{\bf{r}}} \right)\bar{\bf{ M}}_{\ell,m}^ > \left( {k_0{\bf{r'}}} \right)} \,,
    &\\
      &{\mathord{\buildrel{\lower3pt\hbox{$\scriptscriptstyle\leftrightarrow$}}\over
 G}_{TM}}\left( {k_0,{\bf{r}},{\bf{r'}}} \right) = \sum\limits_{\ell,m} {{\tilde c_\ell }{\bf{N}}_{\ell,m}^ < \left( {{k_1}{\bf{r}}} \right)\bar{\bf{ N}}_{\ell,m}^ > \left( {k_0{\bf{r'}}} \right)} \,,
        &
      \end{aligned}
\end{equation}
where $ {k_1} = {n_r}k_0 = \sqrt {{\varepsilon _1}} k_0 $. substituting Eq.(\ref{TE+TMGF}) into Eq.~(\ref{eq:fullGF}), we obtain the expansion coefficients as \cite{res,BH,SM}
\begin{eqnarray}
{c_\ell }&=&  {{ - {k_0}/x} \over {  \left[ { x {h'}_\ell ^{(1)}   {(x) {j_\ell }({x_1})}   { - {x_1}h_\ell ^{(1)}(x)}  {j'_\ell }({x_1})} \right] }}   \equiv {{ - {k_0}/x} \over {{D_\ell }({k_0})}}\\
{\tilde c_\ell } &=& {{ - {k_0}/x} \over { n_r[{x}h_\ell ^{(1)}\left( {{x}} \right)]'{j_\ell }\left( {{x_1}} \right) -n_r^{-1} h_\ell ^{(1)}\left( {{x}} \right)[{x_1 }{j_\ell }\left( {{x _1}} \right)]'}}
\end{eqnarray}
where  $x=k_0R$, $x_1=k_1R$, and $R$  is the radius of MS.

%\section{Emission spectra}
The spontaneous emission rate of an emitter, $R_{sp,i}$ at position ${\bf r}_j$ is proportional to the photon DOS. The rigorous definition of the Purcell factor is  the ratio of photon DOS inside the resonator to the photon DOS of the free space with dielectric constant $\varepsilon_1$, which can be expressed analytically as $P_F(k_0,{\bf r}_j)=\hat\alpha \cdot Im \mathord{\buildrel{\lower3pt\hbox{$\scriptscriptstyle\leftrightarrow$}}\over G} \cdot \hat\alpha/\hat\alpha \cdot Im \mathord{\buildrel{\lower3pt\hbox{$\scriptscriptstyle\leftrightarrow$}}\over G}_1 \cdot \hat\alpha \equiv P^{TE}_F(k_0,{\bf r}_j)+P^{TM}_F(k_0,{\bf r}_j)$, where $\hat\alpha$ denotes the polarization direction of the dipole source. $\mathord{\buildrel{\lower3pt\hbox{$\scriptscriptstyle\leftrightarrow$}}\over G}_1 $ is ${\mathord{\buildrel{\lower3pt\hbox{$\scriptscriptstyle\leftrightarrow$}}\over G}_0}$ given in Eq.~(3) with $k_0$ replaced by $k_1$. Solving the Dyson equation (2) yields\cite{SM}
\begin{equation} P^{TE}_F (k_0,{\bf r}_j)=\sum_{\ell,m} Im \left( \frac {2\pi N_{\ell}(k_0)} {D_{\ell}(k_0)}\right) |j_{\ell}(k_1r_j)|^2, \end{equation}
where $N_{\ell}(k_0)$ is related to $D_{\ell}(k_0)$ in Eq.~(5) by swapping the spherical Bessel function $j_{\ell}(x_1)$ with the spherical Neumann function $y_{\ell}(x_1)$. Similar expression holds for $P^{TM}_F(k_0,{\bf r}_j)$\cite{SM}.

For ZnO MSs considered here,  the light emission which spreads over the entire visible range is related to defect states\cite{defects}.
We consider that under constant optical pumping, the dipole sources related to defect states inside the ZnO microsphere are evenly distributed. Each light source emits photons incoherently. In such case, we can simply calculate the EM fields $({\bold E}_i,{\bold H}_i)$ produced by each light source $i$, and its emission spectrum can be obtained by the Poynting factor ${\bold P}_i={\bold E}_i\times{\bold H}_i^*$ of the EM field at point ${\bf r}$ outside the microsphere. The total emission spectrum is the sum of the contributions from all light sources inside the microsphere.\

Including the Purcell factor $P_F$ we can write
\begin{eqnarray} \label{ITE}
 {I_{sp}^{TE}} &=& \frac 1 V \int_{ {\rm{ f.p.}}}d{{{\bf r}'}}  {\int d{\bf r}_j P_{F}(k_0,{\bf r}_j)  {{\mathop{\rm Re}\nolimits} \left\{ {{{\hat e}_z} \cdot {\bold P}_{TE,j}} \right\}} } \nonumber \\
 &=&  B_{sp}(k_0)\sum_{\ell,m}S_{\ell}(k_0) |c_{\ell}|^2 {\cal F}_{\ell m}(k_0),
\end{eqnarray}
where $B_{sp}(k_0)=\sum_i \delta(\omega_i-ck_0) R_{sp}(\omega_i)/P_F(\omega_i) $ denotes the source distribution, which is related to DOS of emitters at frequency $\omega=ck_0$ weighted by the population distribution. "f.p." stands for "focal plane". $S_{\ell}(k_0)=\frac 1 V \int P_{F}(k_0,r_j) |j_{\ell}(r_j)|^2 r_j^2 dr_j$ is the volume-averaged Purcell factor, which is the source term in channel $\ell$.
%\begin{equation}\label{Far}
%{\cal F}_{\ell m}(k_0)= \int_{\rm{ fp}} d\bf{r'}{\mathop{\rm Re}\nolimits} \left\{ {\hat e}_z \cdot \left( -i \bar{\bf{M}}_{\ell,m}^> ( k_0{\bf{r'}} %)\right) \times \bar{\bf{N}}_{\ell,m}^{> *}( k_0\bf{r'}) \right\}.
%\end{equation}
$|c_{\ell}|^2{\cal F}_{\ell m}(k_0)$ has the physical meaning as the leak-out probability\cite{SM}, which is weaker for resonant modes of higher $Q$, since these modes become better confined. The counterpart for the TM mode reads as Eq.~(\ref{ITE}) but with  $c_l$ and ${\bf{M}}_{\ell,m}^ < \left( {k_1{\bf{r} }_j} \right)$ replaced by $\tilde c_\ell $ and ${\bf{N}}_{\ell,m}^< \left( {k_1{\bf{r} }_j} \right)$.

We now describe a  scheme to separate out the contributions from leaky modes and resonant modes. For leaky modes, the Purcell factor, $P_F$ in Eq.~(\ref{ITE}) is replaced by a channel-averaged Purcell factor $\bar P_F(k_0)$, excluding the contribution from resonant modes, which becomes nearly constant with an average value around 1.2, 0.11, and, 0.07, respectively for the 1.8, 3.7, and 5 $\mu m$ MSs\cite{SM}. This contribution can be directly calculated from the full GF of the micro cavity multiplied by $\bar P_F(k_0)$, which we denote as $I_{leak}$. The results for three ZnO MSs are shown as orange curves in  part (a) of Figs. 1-3. Mode numbers are labeled by $n/\ell$ with TE (TM) modes in black (red), where $n$ denotes the index of roots for a give angular mode $\ell$ with smaller $n$ corresponding to longer wavelength and higher $Q$. For small-size microspheres, $I_{leak}$ is rather weak as seen in Fig.~1(a), since the Purcell factor is inversely proportional to cavity volume and only modes with very small $Q$ can be considered as leaky modes. For larger MSs, $I_{leak}$ gives rise to a broad background emission profile with weak resonance features as seen in Figs. 2(a)  and 3(a). These weak structures are actually stronger at higher order modes ($n>1$) than at principal modes ($n=1$), since their corresponding leak-out probability is higher.

\begin{figure}[t!]
\centering
\includegraphics[trim=2.5cm 1.5cm 3cm 0cm,clip,width=3.5in]{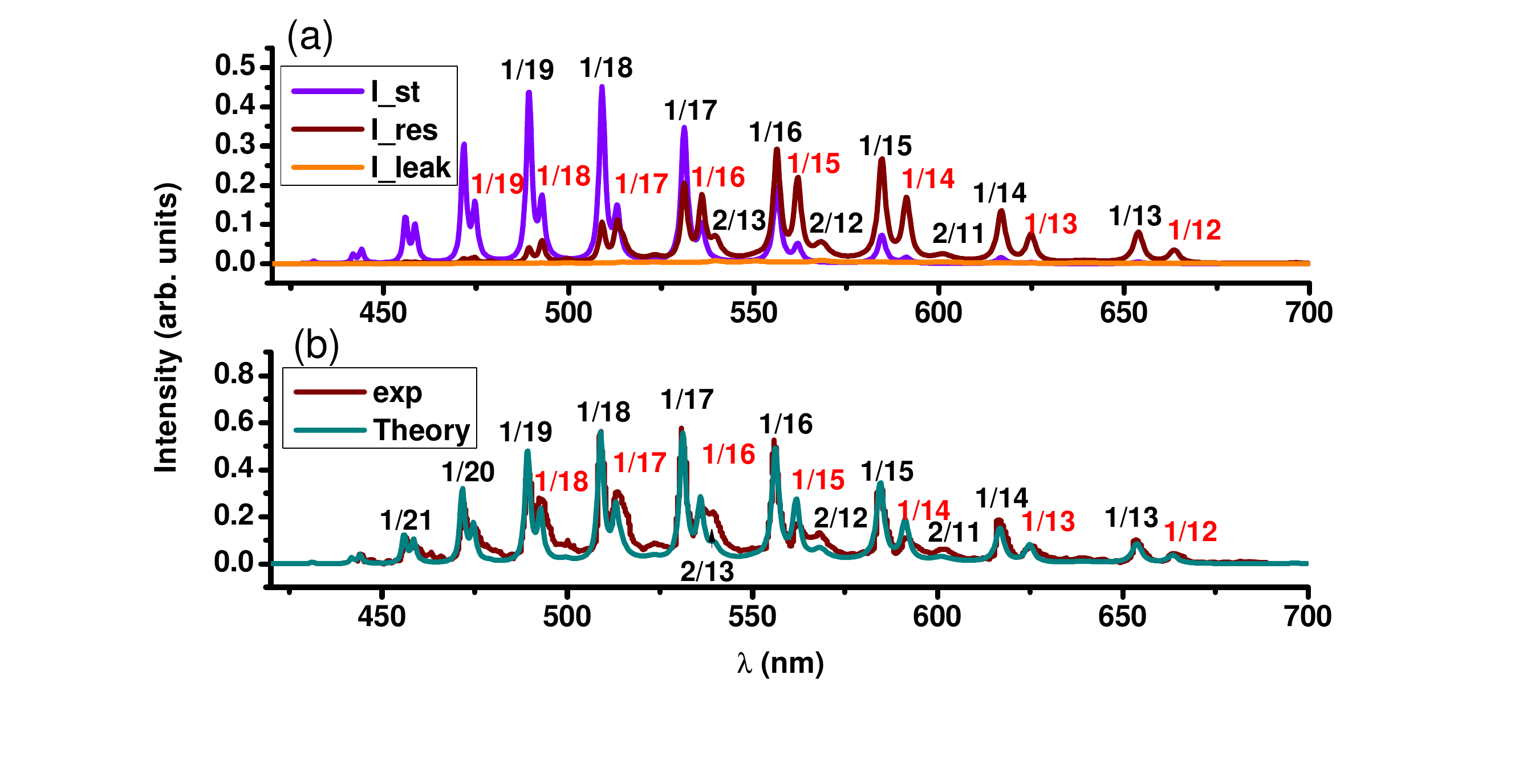}
   \caption{\small (a) Contributions due to leaky modes (orange), resonant modes in spontaneous emission  (brown), and stimulated emission (purple) for a ZnO microsphere of diameter 1.8$\mu m$ with a small porosity at 0.018.  (b)PL spectrum (brown) taken from Ref. \onlinecite{abl} and theoretically predicted emission spectrum (cyan) by using $Q_{ex}=3\times 10^{7}$ and $f_r=3\times 10^{-6}$.}
   \label{fig:1}
\end{figure}

The resonant-mode contribution ($I^{TE}_{res}$) is described by the difference $I_{sp}-I_{leak}$, which is proportional to the  Purcell factor at the resonance frequency. Mathematically, we can write the function $1/{D_\ell \left( {k_0} \right)}$ as a sum over its poles in the complex $k_0$ plane as\cite{SM}
\begin{equation}\label{pole}
\frac 1  {D_\ell \left( k_0 \right)} =  \sum_n \frac {1} {{D_\ell }'(k_{n\ell}) (k_0 - k_{n\ell}) },
\end{equation}
where $k_{n\ell}$ denotes the complex roots of the transcendental equation $D_\ell \left( {k_0} \right)=0$ and ${D_\ell }'$ denotes the derivative of ${D_\ell}$.
The expansion described in Eq.~(\ref{pole}) is rigorous if all poles are included. For  $I_{res}$ the expansion of $1/{D_\ell \left( {k_0} \right)}$ can be expressed as a sum  only high-$Q$ modes, while contributions of all low-$Q$ modes (whose accurate roots are difficult to find) are included in $I_{leak}$.  The full GF of a cavity of arbitrary shape can also be solved numerically via an expansion of quasi-normal modes (QNMs)\cite{15,16}.  For a spherical cavity, we show that the expansion coefficients can be derived analytically and precisely, and the approach used here does not need to worry about the normalization problem of QNM wavefunctions, as the full GF used is already properly normalized.

Consequently, we obtain
\begin{equation}
I^{TE}_{res}(k_0)= B_{sp}(k_0) \sum\limits_{n,\ell }\frac {  (S_{n\ell}-S^0_{n\ell}\bar P_F ) {\cal F}_{\ell}(k_0)}{ \left| R {D_\ell }'\left( k_0 \right) \left( k_0 - k_{n\ell} \right) \right|^2} ,
\end{equation}
where $S_{n\ell}=S_{\ell}(k_{n\ell})$ and  $S^0_{n\ell}=\int |j_{\ell}(r_j)|^2 r_j^2 dr_j/(\mu m)^3$.

$I^{TE}_{res}$ as a function of wavelength is shown as a brown curve in part (a) of Figs. 1-3. For the 1.8 $\mu m$ MS, the  TE and TM principal modes with mode number $\ell$ and $\ell-1$, respectively (the peaks labeled $1/\ell$ in black and $1/(\ell-1)$ in red in Fig. 1(a)) are close to each other  with  the TE peak  stronger than the TM peak for $\lambda > 550$ nm, but  the trend is reversed for  $\lambda < 550$ nm, although the $Q$ in TE mode is higher. This indicates a competing mechanism between the Purcell factor and the leak-out probability. The same trend is found in the 3.7 $\mu m$ MS with the reversed trend occurring at $\lambda \approx 650$ nm. For the 5 $\mu m$ sphere, the TE and TM modes resonant frequencies are very close, so it is hard to tell from the figure.

\begin{figure}[t!]
\centering
\includegraphics[trim=2.5cm 1.5cm 3cm 0cm,clip,width=3.5in]{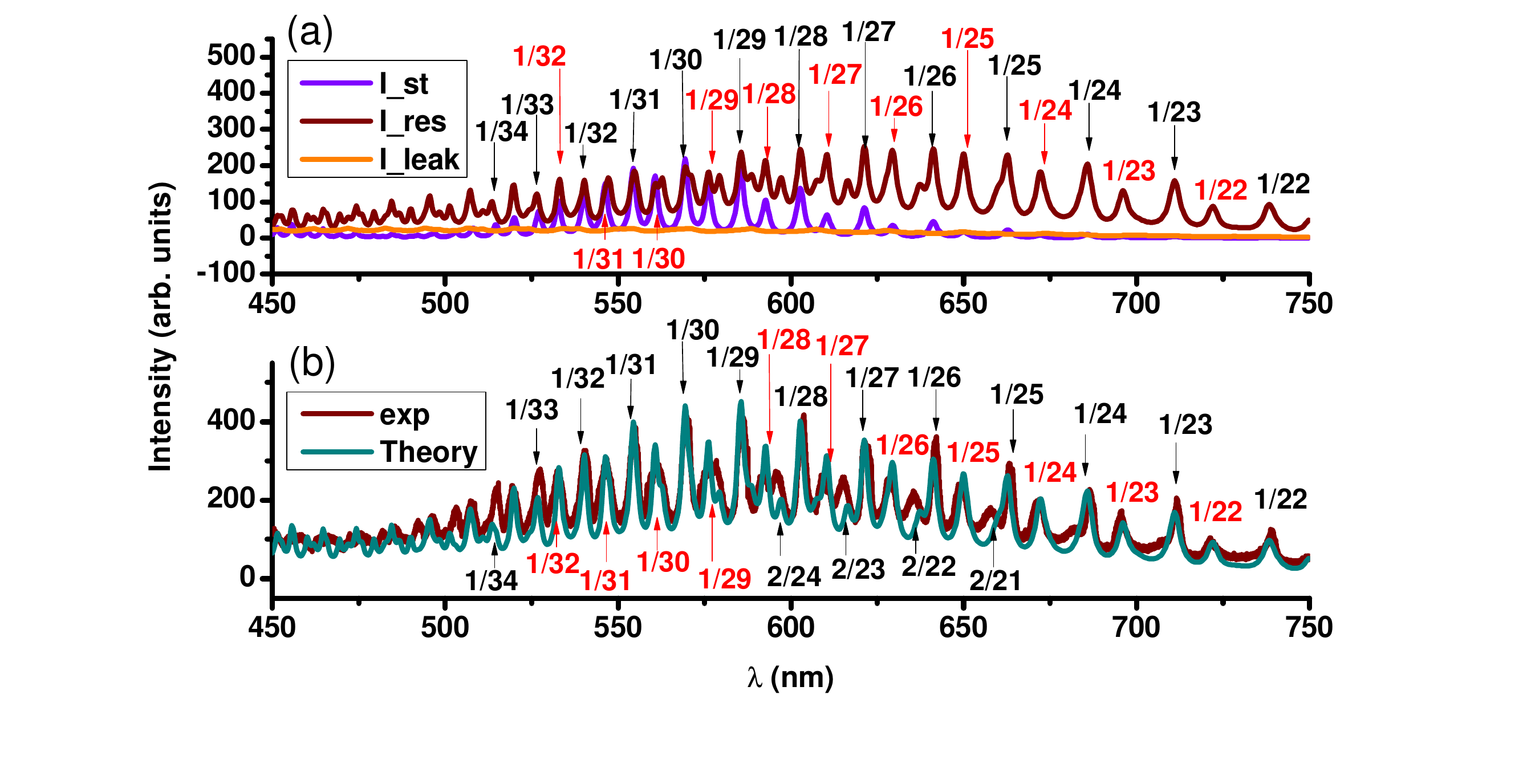}
   \caption{\small (a) Contributions due to leaky modes (orange), resonant modes in spontaneous emission  (brown), and stimulated emission (purple) for a ZnO microsphere of diameter  3.7$\mu m$ with a large porosity around 0.28.
(b) PL  spectrum taken from Ref.~\onlinecite{Trong} (brown) and theoretically predicted emission spectrum (cyan) by using $Q_{ex}=2 \times 10^{7}$ and $f_r=2\times 10^{-7}$. (See SM for details)\cite{SM} }
   \label{fig:2}
\end{figure}

Finally,  the resonant-mode contribution to the stimulated emission intensity is given by $I_{sp}$ multiplied by $ Q_{n\ell}$ for each resonant-mode, and we have

\begin{equation}
I^{TE}_{st}(k_0) = B_{st}(k_0) \sum\limits_{n,\ell }\frac{S_{n\ell}Q_{n\ell} {\cal F}_{\ell}(k_0)}  { \left|R {D_\ell }'\left( k_0 \right) \left( k_0 - k_{n\ell} \right) \right|^2 }  \, ,
\end{equation}
where $B_{st}(k_0)=\sum_i \delta(\omega_i-ck_0)R_{st}(\omega_i)/Q_i P_F(\omega_i) $. In the PL emission of ZnO microspheres, the upper levels of the emitters are states near the bottom of the conduction band which are occupied by electrons relaxed after photoexcitation, while the lower levels correspond to defect states distributed near the mid gap\cite{defects} and are partially depopulated under photoexcitation. Thus, the ratio of source distribution function for stimulated and spontaneous emission $B_{st}(k_0)/B_{sp}(k_0)$ can be approximated by a constant $f_r$, which is proportional to the parameter $g_i \gamma_i \tau_0/(1-\tau_i \gamma_i  g_i)$ averaged over the emitters and depends on the pump power of the excitation source.
The  stimulated emission intensity $I_{st}(k_0)$ as  a function of wavelength is shown as purple curves in part (a) of Figs. 1-3. It is found that $I_{st}$ becomes dominant for short wavelengths, where the $Q^2$ factor overcomes the weak leaking before it reaches $Q_{ex}$. Due to the constraint of $Q_{ex}$, $I_{st}$ typically reaches a maximum near the "saturation" point where $Q_{0,n\ell}\approx Q_{ex}$.
For $I_{st}$ spectrum, the TE mode peak is always stronger than the corresponding TM mode peak due the  strong enhancement in $Q^2$ factor, which overcomes the weak leak-out probability.

\begin{figure}[t!]
\centering
\includegraphics[trim=1cm 0cm 1cm 0cm,clip,width=3.5in]{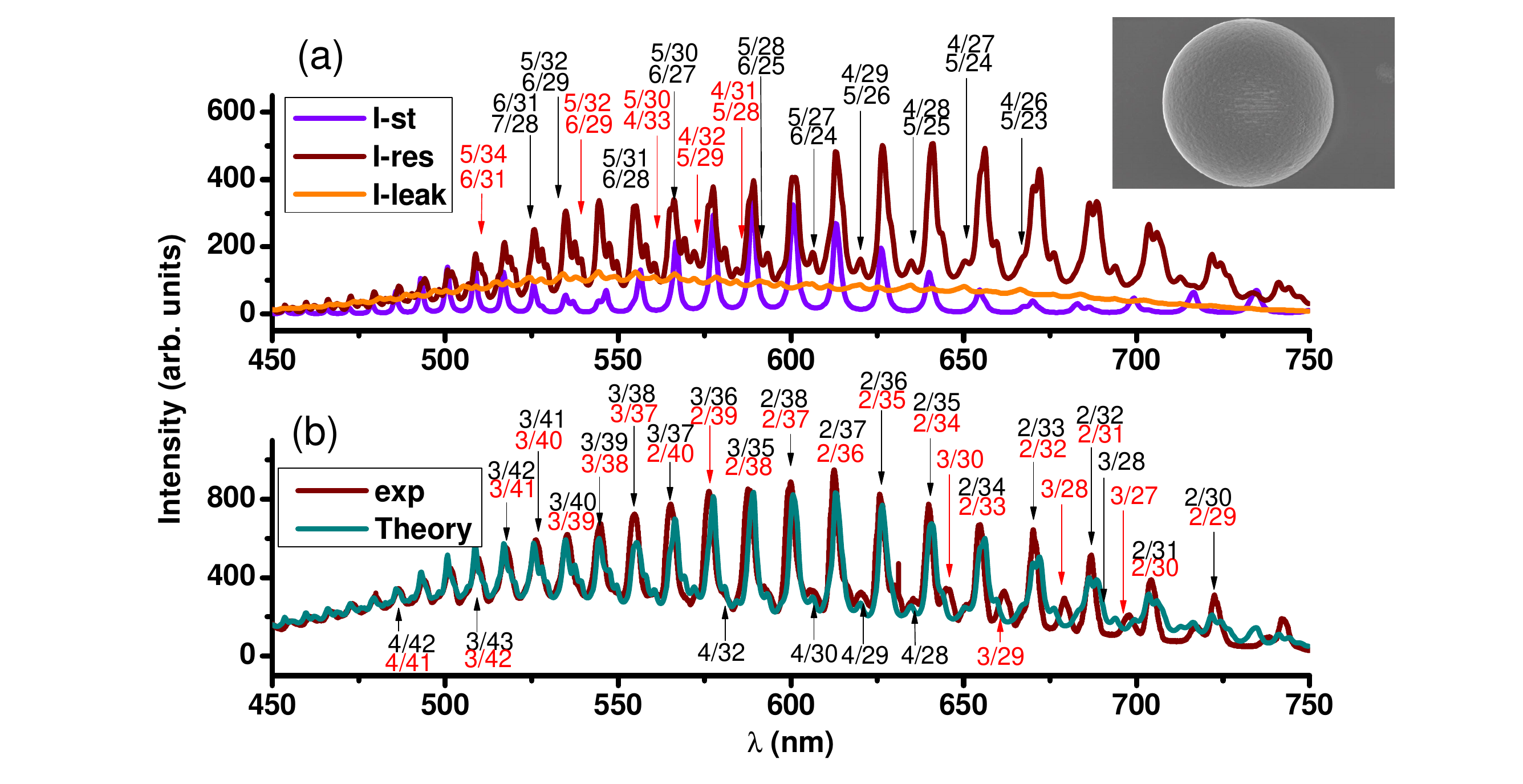}
   \caption{\small (a) Contributions due to leaky modes (orange), resonant modes in spontaneous emission  (brown), and stimulated emission (purple) for a ZnO microsphere of diameter 5.04$\mu m$ with a porosity around 0.14. The inset shows the SEM picture of the ZnO MS fabricated for this work.
(b) Experimental data of PL spectrum obtained (brown) and  theoretically predicted emission spectrum (cyan) by using $Q_{ex}=2\times 10^{8}$ and $f_r=2\times 10^{-8}$. (See SM for more details)\cite{SM} }
   \label{fig:3}
\end{figure}

In part (b) of Figs. 1-3, we compare our theoretical results for $I_{sp}+I_{st}$ on three samples of ZnO microsphere synthesized under different conditions. The sample of Fig. 1 is fabricated in superfluid helium by laser ablation to achieve high-Quality single crystal with smooth surface\cite{abl}, while the sample from Figs. 2 and 3 are made by chemical synthesis, which leads to more crystal imperfection in the form of air holes. The effective medium theory\cite{medium} allows us to estimate the effect of porosity for these samples. By taking into account the finite porosity, Purcell effect and the contribution of stimulated emission under continuous laser pumping, the calculated ratio of the total intensity between the TE and TM modes of various orders (for $\ell$ up 40 to and $n$ up to 4) all match the experimental data well. Note that the Q factor for each WGM is determined by the poles of GF, not a fitting  parameter put in phenomenologically.  The porosity value used is dictated by the spacing between dominant TE and TM modes, which are not necessarily the principal modes. In fact, for the 5 $\mu m$ MS (and likely for larger MSs), the PL spectrum is dominated by higher-order modes ($n$=2 and 3). The only fitting parameters used to determine relative intensities of all resonant peaks are $Q_{ex}$ and $f_r$, once the source distribution $B_{sp}(k_0)$ is determined. Our model captures quantitatively the salient features of  principal  and higher-order TE and TM modes for all three MSs. For the $3.7\mu m$ and $5\mu m$ MSs, relative peak intensities of more than 30  peaks are well accounted for. Even tiny fine features such as the $n=4$ modes for $\ell=28-32$ for the $5\mu m$ MS are observed experimentally and identified theoretically with good agreement in lineshape. Strikingly, the principal modes have negligible contribution in the visible range for the $5\mu m$ MS, which is mainly caused by $Q_{ex}$\cite{SM}.

In conclusion, we demonstrated an useful scheme to simulate the PL spectrum of ZnO MSs and obtained excellent fit for MSs of various sizes. Through careful comparison of theory and experiment, we learned the roles played by stimulated emission, Purcell factor, and leaky modes. This also allows one to understand the effect of porosity in chemically synthesized MSs and the limitation on PL emission caused by $Q_{ex}$. All these information are valuable for the design of MS cavities for applications in light-emitting and optical sensing devices.

\begin{acknowledgments}
%\section*{Acknowledgments}

Work supported in part by Ministry of Science and Technology (MOST), Taiwan  under contract nos. 106-2112-M-001-022 and 107-2112-M-001-032. We thank S. Y. Shiau and S. W. Chang for fruitful discussions and P. J. Cheng for assistance in making figures.

\end{acknowledgments}

%\section{Comparison with experiment}

% $Q_{ex}\approx 10^-8$ used is reasonable\cite{Qex}

%
%%%%%%%%%%%%%%%%%%%%%%%%%%%%%%%%%%%%%%%%%%%%%%%%%%%%%%%

\end{document}